\documentclass[aps,prb,twocolumn,amsmath,amssymb,gallery,footinbib,superscriptaddress,floatfix,nobibnotes]{revtex4-2}
\usepackage{graphicx}
\usepackage{dcolumn}
\usepackage{bm}
\usepackage{hyperref}
\usepackage{stackrel,amssymb}
\hypersetup{colorlinks=true, linkcolor=blue, citecolor=blue, urlcolor=magenta, pdftitle={On the search of displacive chiral phase transitions}, pdfauthor={F. Gomez-Ortiz and E. Bousquet}}
\usepackage{fixltx2e}
\usepackage{bbding}
\usepackage{xcolor} 

\begin{document}
\preprint{APS/123-QED}
\title{Switchable Skyrmion--Antiskyrmion Tubes in Rhombohedral BaTiO$_\mathrm{3}$ and Related Materials}
\author{Fernando G\'{o}mez-Ortiz}
\thanks{These authors contributed equally}
\email{fgomez@uliege.be} 
\affiliation{Theoretical Materials Physics, Q-MAT, Université de Liège, B-4000 Sart-Tilman, Belgium}
\author{Louis Bastogne}
\thanks{These authors contributed equally} 
\affiliation{Theoretical Materials Physics, Q-MAT, Université de Liège, B-4000 Sart-Tilman, Belgium}
\author{Sriram Anand}
\affiliation{Theoretical Materials Physics, Q-MAT, Université de Liège, B-4000 Sart-Tilman, Belgium}
\author{Miao Yu}
\affiliation{Theoretical Materials Physics, Q-MAT, Université de Liège, B-4000 Sart-Tilman, Belgium}
\author{Xu He}
\affiliation{Theoretical Materials Physics, Q-MAT, Université de Liège, B-4000 Sart-Tilman, Belgium}
\author{Philippe Ghosez}
\email{philippe.ghosez@uliege.be}
\affiliation{Theoretical Materials Physics, Q-MAT, Université de Liège, B-4000 Sart-Tilman, Belgium}
\date{\today}
\begin{abstract}
Skyrmions are stable topological textures that have garnered substantial attention within the ferroelectric community for their exotic functional properties. While previous studies have questioned the feasibility of [001]$_{\text{pc}}$ skyrmion tubes in rhombohedral BaTiO$_3$ due to the high energy cost of 180$^\circ$ domain walls, we demonstrate here their stabilization with topological charges of $\mathcal{Q} = \pm 1$ from density functional theory and second-principles calculations. By enabling extensive vortex and antivortex polarization configurations, the expected prohibitive energetic barriers are overcomed while preserving the topological nature of the structures. Notably, we extend these findings to demonstrate the appearance of skyrmion and antiskyrmion tubes in other related materials, highlighting their broader relevance. Furthermore, our computational experiments indicate that these structures can be directly stabilized and reversibly switched by applied electric fields, establishing a straightforward route for their practical realization and functional control in nanoelectronic devices.
\end{abstract}
\maketitle
Skyrmions have garnered significant interest in condensed matter physics due to their unique topological properties, which impart remarkable stability against perturbations~\cite{Nagaosa-13}.
First observed as two-dimensional magnetic textures~\cite{Muhlbauer-09}, skyrmions can extend into three dimensions as cylindrical or tubular structures called skyrmion tubes, which maintain translational invariance along a specific axis~\cite{Wolf-22}.

Analogous nanoscale polarization textures have been predicted~\cite{Nahas-15,Mauro-19,Mauro-24} and observed~\cite{Das-19} in ferroelectric materials, (although it was suggested they might not be ubiquitous~\cite{Zatterin-24}), drawing attention due to their smaller size~\cite{Mauro-19} and fascinating functional properties including negative capacitance~\cite{Iniguez-19,Das-21}, emergent chirality~\cite{Louis-12,Shafer-18,Kim-22} or ultrafast dynamics~\cite{Zhu-22,Aramberri-24,Prokhorenko-24}. 

Despite experimental efforts, ferroelectric skyrmion tubes have so far only been experimentally observed in superlattices, while their emergence in bulk materials remains elusive. Among potential candidates, PbTiO$_3$ was theoretically proposed~\cite{Mauro-19} owing to its capacity to host non-collinear polar arrangements within its domain walls~\cite{Wojdel-14}, yet experimental confirmation of their topological nature remains unachieved.

In contrast, recent findings predicted fractional Skyrme lines in rhombohedral BaTiO$_3$ confined within 180$^\circ$ domain walls~\cite{Halcrow-24}. However, the authors were unable to stabilize ordinary skyrmions.
Interestingly, antiskyrmion tubes with $\mathcal{Q}=-2$ have been predicted in BaTiO$_3$ when the columnar nanodomain aligns with the polarization easy axis in any $\langle111\rangle_{\rm{pc}}$ direction broadening the scope of topological structures in this material~\cite{Mauro-24}. Besides, meron-antimeron lattices have also been stabilized by means of acoustic phonon excitations~\cite{Bastogne-24}.
Adding to this promising landscape, recent experiments have shown vortex-antivortex lattices in twisted BaTiO$_3$ freestanding layers~\cite{Santolino-24} and center-convergent polar domains in BaTiO$_3$ nanoislands~\cite{Olaniyan-24}. Amazingly, despite ferroelectricity in BaTiO$_3$ being first reported almost 80 years ago~\cite{Merz-49}, new discoveries continue to emerge recurrently. These findings position BaTiO$_3$ as a promising candidate for skyrmion observation in bulk ferroelectrics.

In this work, combining density functional theory and second-principles~\cite{wojdel2013first,escorihuela2017efficient} calculations, we report the stabilization of translationally invariant polarization textures along the $\left[001\right]_{\rm{pc}}$ direction in BaTiO$_3$ characterized by skyrmion numbers of $\mathcal{Q}=\pm 1$. Computational experiments demonstrate the stabilization and controllable switching of these textures, offering a clear route to their practical realization.
Notably, this is the first demonstration of skyrmions and antiskyrmions in the same ferroelectric system under identical conditions.
Additionally, we generalize our findings to other related materials such as KNbO$_3$. 
Our results challenge recent interpretations~\cite{Halcrow-24,Mauro-24}, which rejected the existence of these type of skyrmion tubes in rhombohedral ferroelectrics due to the prohibitive energy cost of the resultant $180^\circ$ domain-walls.
Here we show that if the polarization in the matrix is allowed to form a vortex- or antivortex-like texture, the energetic constraints are alleviated, enabling their stabilization.
\section{Results}
\label{sec:Results}

\begin{figure}[tb]
     \centering
        \includegraphics[width=\columnwidth]{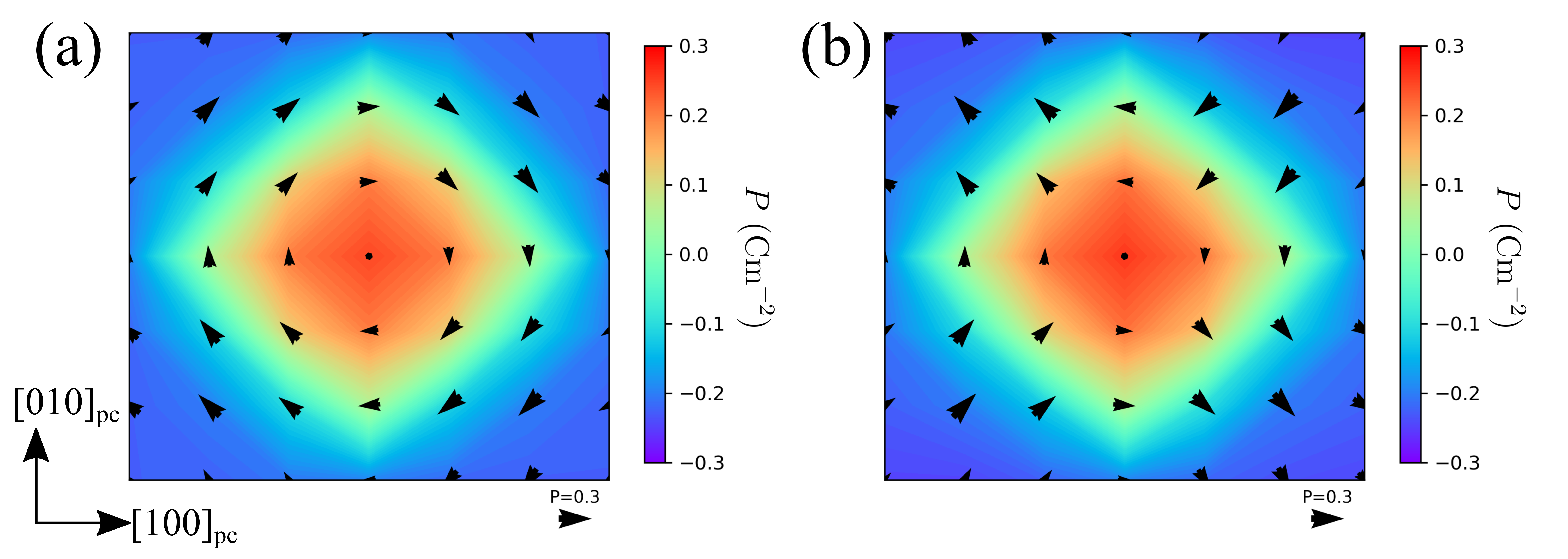}
      \caption{(a) DFT equilibrium skyrmion texture obtained after relaxation of manually imposed initial Ti displacements. (b) Equilibrium antiskyrmion texture obtained after relaxation of the skyrmion texture after reverting the sign of the displacements along $x$. Arrows indicate the in-plane components of the polarization, whereas the color map represents the out-of-plane polarization computed as explained in Supp. Inf. the supercell size used is 7x7x1.}
      \label{fig:Figure1} 
  \end{figure}

\noindent\textbf{$\left[001\right]_{\rm{pc}}$-oriented skyrmion tubes}
In light of the computational experiment presented in Ref.~\cite{Mauro-19} where a columnar nanodomain was embedded in a matrix of antiparallel polarization within tetragonal PbTiO$_3$, it is natural to wonder whether similar nanocolumns can be stabilized in rhombohedral ferroelectrics such as BaTiO$_3$ or KNbO$_3$.
Building on this concept, we construct a $\left[001\right]_{\rm{pc}}$-oriented nanocolumn in rhombohedral BaTiO$_3$ where the Ti displacements responsible for the in-plane and out-of-plane polarization components are intentionally adjusted to achieve the desired configuration. The DFT equilibrium structure obtained after relaxation (details in Supp. Inf.) is presented in Fig.\ref{fig:Figure1}(a).

Notably, in PbTiO$_3$ skyrmions~\cite{Mauro-19}, the $\left[111\right]_{\rm{pc}}$-oriented antiskyrmion tube in BaTiO$_3$~\cite{Mauro-24} and the Skyrme lines appearing in $180^\circ$ domains in BaTiO$_3$~\cite{Halcrow-24} the non-colinear character of the polarization is confined within the domain wall. In contrast, our results demonstrate that the Bloch-character of the domain extends throughout the matrix of antiparallel polarization after relaxation.
As shown in Fig.~\ref{fig:Figure1}(a), even though the out-of-plane polar nanocolumn is confined to few unit cells, the vortex-like disturbance of the field, reflected by the in-plane components of the polarization, extends throughout the whole supercell. 
This behavior can be attributed to two main factors: (i) the development of an in-plane polarization component reduces elastic energy, allowing each cell to locally adopt a configuration closer to the nearest $\langle111\rangle_{\rm{pc}}$ R3m ground state, and (ii) it mitigates the formation of energetically unfavorable $180^\circ$ domains, which would otherwise be prohibitively costly, as reported in Refs.~\cite{Mauro-24,Halcrow-24}. 

Interestingly, a similar behavior is reported in Ref.~\cite{Mauro-19}, where PbTiO$_3$ skyrmion tubes exhibit a comparable response when subject to an expansive tensile strain exceeding a critical threshold. 
This observation suggests that such behavior could, in fact, be more general and extendable to other rhombohedral ferroelectrics. Notably, we demonstrate here that similar structures can also be stabilized in KNbO$_3$ (see Fig.~S1, Fig.~S2) further reinforcing this idea. 

Importantly, even though the continuous deformation from a homogeneous background to a vortex-like polarization configuration in the matrix helps stabilize the structure, it does not alter its topological properties. The non-trivial topology of a skyrmion tube can be computed by integrating the Pontryagin density (see supplementary information).
As shown in Fig.~S3(a), an integer value of $\mathcal{Q}=1$ is obtained for the polar texture schematized in Fig.~\ref{fig:Figure1}(a), proving its non-trivial character.

Unlike the antiskyrmions observed in Ref.~\cite{Mauro-24}, the skyrmions reported here exhibit a chiral nature. The existence of chiral behavior can be experimentally probed using resonant soft X-ray diffraction–based circular dichroism measurements~\cite{Louis-12,Shafer-18,Lovesey-18,Chauleau-20,McCarter-22}, and is a property of significant relevance for light-matter interactions, with potential applications in optoelectronics~\cite{Shen-24}.

Computationally, the chirality can be computed by means of the helicity~\cite{Shafer-18,Junquera-23,Moffat-92,Moffatt-14}. As shown in Fig.~S3(c), the chiral nature of the skyrmion tube is determined by the coupling between the in-plane polarization rotation and the polarization value at the core, granting a non-zero net value.

Due to the periodic boundary conditions and the Poincaré-Hopf theorem~\cite{Milnor-65}, the total vorticity in the simulation supercell must sum to zero. Consequently, the vortex and antivortex textures must counterbalance each other. As it can be observed in Fig.~\ref{fig:Figure1}(a), the antivortex is located at the middle of the cell edges. One may wonder whether the center of the nanocolumn could be positioned at the antivortex core to stabilize an antiskyrmion with a topological charge of $\mathcal{Q}=-1$ in place of a skyrmion. This can be accomplished by reversing the sign of the atomic displacements along the $x$-direction of the skyrmion texture with respect to the cubic reference, effectively transforming vortices into antivortices and vice versa.

Interestingly, and in contrast to tetragonal PbTiO$_3$~\cite{Mauro-19},  we find that the position of the nanocolumn is flexible and decoupled from the in-plane components, allowing it to align with the antivortex core and stabilizing an antiskyrmion, as illustrated in Fig.~\ref{fig:Figure1}(b). The DFT energy of the structure is almost degenerate with that of the skyrmion (within the accuracy of DFT, see Fig.S2), demonstrating that rhombohedral ferroelectrics can host both skyrmion and antiskyrmion textures under identical strain and growing conditions.

Finally, we go a step further by revealing the existence of two distinct types of skyrmions and antiskyrmions: one centered at Ti sites and the other at Ba sites.
Although initially stabilized by intentional Ti displacements, resulting in vortex and antivortex cores centered at Ti sites, we found that configurations with the vortex/antivortex singularity centered at Ba sites are also stable. Notably, the Ba-centered configurations are energetically more favorable, lying $1.2$ meV/f.u. lower than their Ti-centered counterparts in our 7$\times$7$\times$1 supercell.
\begin{figure*}[tb]
     \centering
      \includegraphics[width=\textwidth]{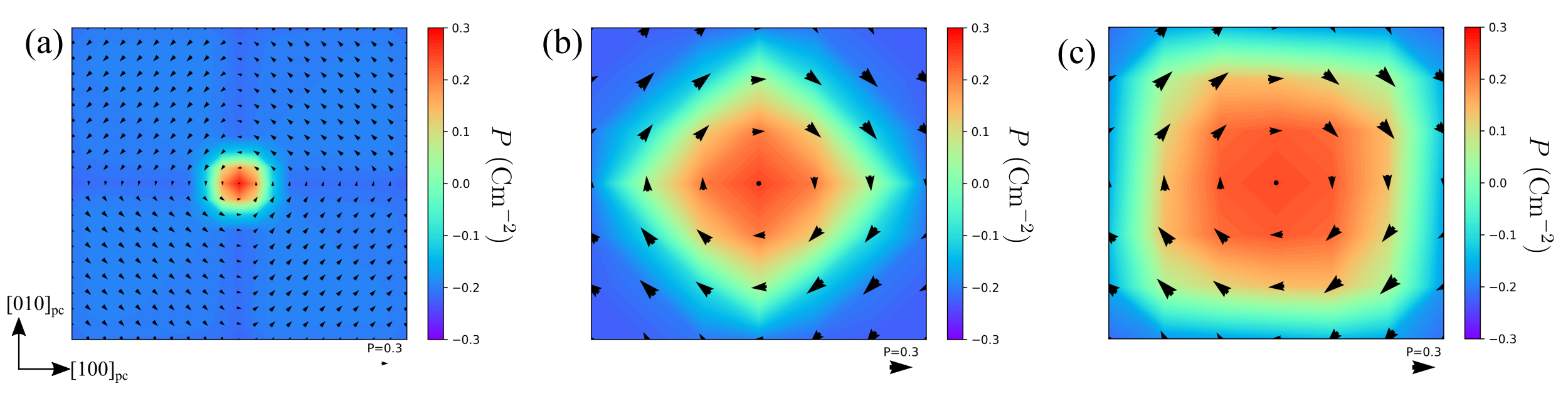}
      \caption{Stabilization of different defects in BaTiO$_3$. (a) $r=1$ u.c. nanocolumn stabilized with the second-principle model. (b) Rhombus and (c) square-like oriented cross sections of the nanocolumn. Arrows indicate the in-plane components of the polarization, whereas the color map represents the out-of-plane polarization computed as explained in Sup. Inf. the supercell sizes used are 21x21x1 and 7x7x1.}
      \label{fig:orientation} 
\end{figure*}
This is explained from the fact that the vortex or antivortex cores represent the intersection of the two domain walls between the up/down and right/left in-plane polarization components which are more stable when centered on the Ba site~\cite{Padilla-96}. We report a detailed energy analysis of both types of skyrmion/antiskyrmion pairs together with nudged elastic band (NEB)~\cite{Jonsson-98,henkelman2000improved,sheppard2008optimization} calculations to compute the energy barriers between them in Fig.~S4.

\noindent\textbf{Orientation, size effects and stability of skyrmion tubes}
The energy and stability of skyrmion tubes are analyzed across different nanocolumn orientations and radii as well as various supercell sizes. Given the comparable energetics of the Ba/Ti-centered skyrmion and antiskyrmion textures, we focus our analysis on the Ti-centered skyrmion case for simplicity. 
Orientation effects are examined with DFT, while stability and size effects are studied with second-principles methods (details in Supp. Inf.).

\emph{Energetics with increasing nanocolumn size.-} The impact of nanocolumn size on the defect's energy is explored for a fixed supercell length, $L=21$ u.c.
Interestingly, stable nanocolumns as small as $r=1$ u.c. are found as shown in Fig.~\ref{fig:orientation}(a). This value is smaller than those encountered for in Ref.~\cite{Mauro-24}, where textures with radii smaller than $4$ u.c. decay to a monodomain configuration~\cite{Mauro-24}.
The defect energy scales linearly with the square root of the nanocolumn area, and metastable states exist at different sizes. For instance, doubling the radius to $r=2$ u.c. raises the energy by $\Delta E=0.07$ meV/f.u., highlighting the tunability of nanocolumn size.

\emph{Influence of the orientation of the nanocolumn's cross section.-} As the skyrmion tube is oriented along the $\left[001\right]$ direction, its cross-sections exhibit a cubic shape. This contrasts with the hexagonal-like cross sections reported in Ref.~\cite{Mauro-24}. Square nanocolumns align either rhombically [domain walls along $\lbrace110\rbrace{\rm{pc}}$, Fig.~\ref{fig:orientation}(b)] or cubically [domain walls along $\lbrace100\rbrace_{\rm{pc}}$, see Fig.~\ref{fig:orientation}(c)].
Both configurations are stable with an energy difference of $0.1$~meV/f.u. in our 7$\times$7$\times$1 supercell.  However, attributing this energy difference solely to the domain orientation is challenging, as a change in orientation inevitably results in a change in the area of the nanocolumn which also influences (in a similar magnitude) the total energy as we have just discussed.
What we can conclude anyway, is that the energy landscape associated with varying the orientation of the nanocolumn’s cross-section is extremely flat, allowing for multiple realizations and potential dynamical transitions between them at finite temperatures as we show in the Supplemental Video 1.

\emph{Energetics with increasing supercell size.-} Keeping a fixed value of two unit cells for the nanocolumn, we can increase the value of the supercell in which the defect is embedded \textit{i.e.} $L=7,~11,~13,~15,~17,~19,~21$. As already discussed above, although the nanocolumn size is fixed, the in-plane polarization components extend across the supercell, locally resembling the $\langle111\rangle_{\rm{pc}}$ R3m configuration, leading to an asymptotic energy behavior as shown in Fig.~S5. This allows us to quantify the energy difference between the defect and the rhombohedral monodomain phase in the infinite supercell limit.
As the supercell grows, regions resembling the $\langle111\rangle_{\rm{pc}}$ monodomain expand uniformly, while the defect's energy contribution remains constant. A power-law model, $E=E_d + a/(x+b)$, accurately describes this behavior (see residuals in Fig.~S5).

The asymptotic energy obtained with respect to the cubic reference is $E_d = -24.87 \pm 0.01$ meV/f.u., closly matching that of the rhombohedral monodomain phase ($E_{\rm{R3m}} = -24.94$ meV/f.u.) and lying below the orthorhombic monodomain phase ($E_{\rm Amm2} = -22.29$ meV/f.u.). This result indicates that the defect introduces only a minor, localized energy disturbance, despite inducing a large-scale disruption in the polarization pattern, highlighting the narrow energy window within which these phases coexist.
\begin{figure*}[tb]
     \centering
      \includegraphics[width=\textwidth]{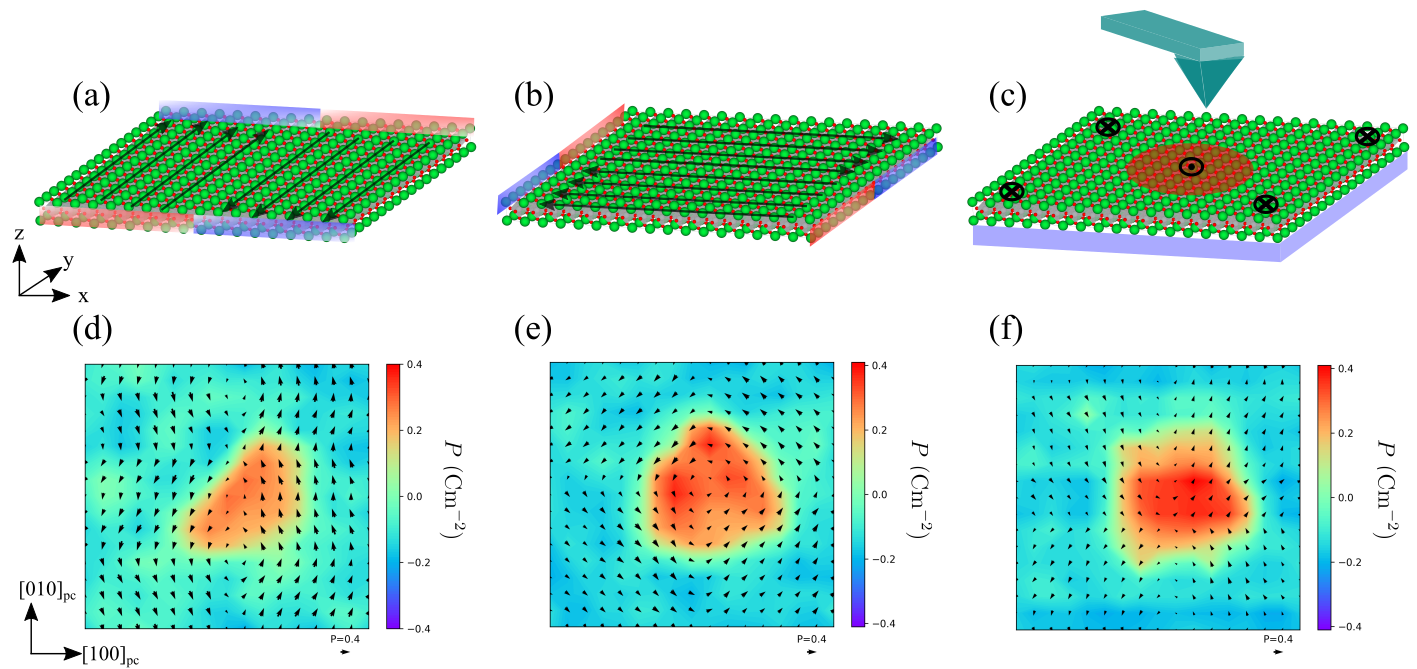}
      \caption{Schematic view of the computational experiment. (a-c) The black arrows denote the desired polarization direction within the material. The red and blue stripes represent electrodes, which can be sequentially activated in a stepwise manner during experimental implementations to achieve the intended polarization (a) Polarization field modulated along the $x$-direction and directed along $y$. (b) Polarization field modulated along $y$ and directed along $x$. (c) Control of the out of plane components achievable by with an AFM tip. (d) Obtained structure after the application of a modulated electric field as described in the main text. (e) Skyrmion texture after in-plane and out of plane components have been controlled with the application of extra fields in configurations (b) and (c). (f) Antiskyrmion texture obtained from (e) after reverting the polarity of the electric field shown in (b). Arrows indicate the in-plane components of the polarization, whereas the color map represents the out-of-plane polarization computed as explained in Sup. Inf.. The supercell size is 17x17x1.}
      \label{fig:comp_exp} 
\end{figure*}

\emph{Stability of the skyrmion tubes.-} Finally, we investigate the stability of skyrmion and antiskyrmions under thermal fluctuations, finding them equally robust. However, Ti centered defects remain stable up to a critical temperature of $T=150$ K, whereas Ba centered defects are stable up to $T=80$ K. Beyond these thresholds, the nanocolumns collapse into a monodomain out-of-plane polarization, while the in-plane vortex components persist. Despite their lower energy, Ba-centered defects exhibit a lower barrier to transitioning into the homogeneous state, as quantified by NEB calculations (Fig.~S6). The transition path is detailed in Supplemental Videos 2 and 3.

\noindent\textbf{Stabilization and switching via applied electric fields}
So far, we have focused on the metastability of skyrmions and antiskyrmions in rhombohedral BaTiO$_3$, where these structures are constructed by intentionally imposing atomic displacements and subsequently relaxing the configurations. In this section, we demonstrate how these textures can be stabilized using inhomogeneous electric fields, also providing an experimental pathway for their realization.

In order to stabilize the in-plane components of the polarization we rely on spatially modulated electric fields. In Fig.~\ref{fig:comp_exp}(a) we schematically show an initial polarization texture. Experimentally, this could be achieved by sequentially activating contiguous electrodes in a stepwise manner, as depicted in the schematic. Numerically, we impose an electric field of $1$ MV/cm modulated along $x$ and directed along $y$ by a cosine function in order to comply with periodic boundary conditions and study the evolution of the system ($L=17$ u.c.) with finite temperature simulations at $T=10$ K. In tetragonal ferroelectrics, such electric fields will result in the stabilization of $180^\circ$ domain walls with the polarization aligned along $y$. However, due to the rhombohedral nature of BaTiO$_3$, in-plane polarization components along the $x$-direction will also develop as well as out-of-plane components resulting on a more complex structure [see Fig.~\ref{fig:comp_exp}(d)].
The direction of the in-plane polarization components along the  $x$ axis can vary and show an erratic behavior as shown in Fig.~\ref{fig:comp_exp}(d) varying depending on the specific realization. However, applying now an electric field oriented along $x$ and modulated along the $y$ direction allows effective control over these in-plane components. This second field induces polarizations along the $x$ axis, similar to those illustrated in Fig.~\ref{fig:comp_exp}(b). Utilizing such configuration will stabilize a vortex, while reversing the polarity of this second field will result in the stabilization of an antivortex texture, as shown in Fig.~\ref{fig:comp_exp}(e-f), respectively. Furthermore, this procedure can be applied recursively; by alternating the polarity of the second electric field, we can achieve a controlled switching between vortex and antivortex states, as demonstrated in Supplemental Video 4. Alternatively, activating simultaneously the electrodes shown in Fig.~\ref{fig:comp_exp}(a) will already stabilize an antivortex texture as shown in Sup. Inf. due to the electric fields generated between contiguous electrodes.
Importantly, due to the nature of the domain-walls only Ba-centered vortex/antivortex can be stabilized in this way.

Similar to the in-plane components, the out-of-plane components will vary depending on the realization when no bias is applied to the sample. To achieve a localized nanocolumn, as shown in Fig.~\ref{fig:comp_exp}(e-f), a Gaussian field as achievable with an AFM tip has been applied on top of a homogeneous background of opposite polarization Fig.~\ref{fig:comp_exp}(c). Experimentally, a homogeneous out-of-plane polarization could first be established via scanning, followed by local reversal to achieve the desired nanocolumn. An important aspect to highlight is that once the nanocolumn is established, it remains stable, and the switching of the in-plane components does not affect it,  in agreement with the DFT results that proved the flexibility of the out-of-plane components to adjust to vortex and antivortex cores. This stability enables the transition from skyrmion to antiskyrmion to occur without the need for an electric field applied along the $z$ direction.
Because the in-plane polarization components extend over a large scale, as discussed throughout the manuscript, highly spatially modulated electric fields are not required, facilitating the experimental implementation of this setup.
\section{Discussion}
\label{sec:Conclusions}
In this study, we have successfully demonstrated, from density functional theory and second-principles calculations, the stabilization of two different types (Ba- and Ti-centered) translationally invariant polarization textures along the $\left[001\right]_{\rm{pc}}$ direction in rhombohedral BaTiO$_3$, characterized by skyrmion numbers of $\mathcal{Q}=\pm 1$, . This constitutes the first theoretical prediction where ferroelectric skyrmions and antiskyrmions are stable in the same material under identical conditions. Our results provide strong evidence that such structures can also be generalized to other rhombohedral ferroelectrics, such as KNbO$_3$ which keeps a rhombohedral phase close to room temperature, broadening the scope of materials capable of hosting complex polar textures and potentially enabling their stabilization under practical operating conditions.

Importantly, our findings challenge the prevailing understanding that $\mathcal{Q}=\pm 1$ skyrmion tubes are energetically unfeasible in rhombohedral ferroelectrics with fixed antiparallel polarization values, where the energy cost of $180^\circ$ domain walls has been considered prohibitive~\cite{Halcrow-24,Mauro-24}. By allowing the polarization matrix to adopt a vortex- or antivortex-like configuration, we demonstrated that these energy constraints can be alleviated while preserving the topological character of the system. Besides, this extended nature of the Bloch-component to the entire matrix can potentially facilitate its experimental observation~\cite{Zatterin-24}.

Additionally, we have delved into various orientations and sizes for the nanocolumn. In contrast to $\left[111\right]_{\rm{pc}}$ BaTiO$_3$ antiskyrmion tubes~\cite{Mauro-24}, cubic cross-sections are found for our skyrmion tubes where rhombus or square orientations are possible for the nanocolumn being essentially degenerate in energy.
The diameter of the nanocolumn is found to be of about $1$ nm, even smaller than the ones reported for $\mathcal{Q}=-2$ antiskyrmions~\cite{Mauro-24}. The energy of the defect in the large scale limit is found to be surprisingly close to the rhombohedral ground state indicating that the defect introduces only a minor and localized energy disturbance. 

Finally, we have demonstrated how these textures can be stabilized and reversibly switched using spatially modulated electric fields, proposing an experimental setup for their realization. Moreover, due to the chiral nature of the skyrmionic state, it can also be experimentally probed by RSXD-CD techniques. To the best of our knowledge, this represents the first system in which the topological charge of a skyrmion can be controlled from $\mathcal{Q}=+1$ to $\mathcal{Q}=-1$ using external electric fields.

This work not only expands the current phase diagram of BaTiO$_3$ but also underscores the potential for stabilizing skyrmion-based polar textures in other ferroelectrics, offering new avenues for the design of materials with tunable topological properties.
\acknowledgments
The authors acknowledge useful discussions with Céline Lichtensteiger. F.G.-O., L.B and Ph. G. acknowledge support by the European Union’s Horizon 2020 research and innovation program under Grant Agreement No. 964931 (TSAR).
TF.G.O. also acknowledges financial support from MSCA-PF 101148906 funded by the European Union and the Fonds de la Recherche Scientifique (FNRS) through the grant FNRS-CR 1.B.227.25F. Ph. G. and X.H. also acknowledges support from the Fonds de la Recherche Scientifique (FNRS) through the PDR project PROMOSPAN (Grant No. T.0107.20).
S. A. acknowledges financial support from the EIT Raw Materials - AMIS Joint Master Program.
\clearpage
\onecolumngrid
\begin{center}
    \textbf{\large Supplementary Material for Switchable Skyrmion–Antiskyrmion Tubes in Rhombohedral BaTiO3 and Related
Materials}
\end{center}
\bigskip 
\section{Methods}
\label{sec:Methods}
 Density functional theory (DFT) calculations were performed using the \textsc{Abinit}~\cite{gonze2020abinit} software package. We used the generalized gradient approximation (GGA) with the PBESol exchange-correlation functional and a plane-wave pseudopotential approach, employing optimized norm-conserving pseudopotentials from the PseudoDojo server~\cite{hamann2013optimized,van2018pseudodojo}. The valence electrons considered were $5s^2 5p^6 6s^2$ for Ba, $3s^23p^63d^24s^2$ for Ti, $2s^22p^4$ for O in BaTiO$_\mathrm{3}$, and $3s^2 3p^6 4s^1$ for K, $4s^24p^64d^45s^1$ for Nb, and $2s^22p^4$ for O in KNbO$_\mathrm{3}$. For both BaTiO$_\mathrm{3}$ and KNbO$_\mathrm{3}$, we used a plane-wave energy cutoff of $40\ \mathrm{Ha}$ and an $8\times 8\times 8$ $\Gamma$-centered k-point mesh, adapting the k-point mesh for the supercell size. The electronic residual self-consistent cycle stopped at a value of $10^{-8}\ \mathrm{Ha}$. 

For BaTiO$_\mathrm{3}$, we carried out Density Functional Perturbation Theory (DFPT) calculations using \textsc{Abinit}~\cite{gonze2020abinit}. Dynamical matrices for the relaxed cubic $Pm\bar{3}m$ structure were computed with an $8\times 8\times 8$ q-point mesh. From the DFPT framework, we also extracted the optical dielectric constant, Born effective charges, strain-phonon coupling and elastic constants.

The Second-principles atomistic models for BaTiO$_\mathrm{3}$ were constructed using \textsc{Multibinit}~\cite{gonze2020abinit} which implements the second-principles approach described in references \cite{wojdel2013first,escorihuela2017efficient}. This method relies on a Taylor expansion of the potential energy surface around the reference cubic structure in terms of all structural degrees of freedom. In our case, we extract the harmonic part directly from the DFPT calculations mentioned above and the anharmonic part is fitted to reproduce the DFT data using the same parameters mentioned above.  

This model, is a slight revision of the one reported in Ref.~\cite{zhang2023structural} including refitted anharmonic terms and additional sixth- and eighth-order terms and has been successfully employed in a previous study~\cite{Bastogne-24}. As noted in the Data Availability section, the model and its validation assessment are publicly accessible.

DFT and second-principles structural relaxations were both performed using the Broyden-Fletcher-Goldfarb-Shanno (BFGS) method~\cite{fletcher2000practical} as implemented in \textsc{Abinit}~\cite{gonze2020abinit} using a $7\times7\times1$ supercell of the 5 atom cubic unit cell at DFT level whereas larger supercells were also considered at the second-principles level. This minimization was performed until the forces are less than $5\times 10^{-5}$ $\rm Ha/Bohr$ and the stresses are less than $5\times 10^{-7}$ $\rm Ha/Bohr^3$. 
Second-principles finite temperature simulations were performed using a Hybrid molecular dynamics Monte-Carlo approach (HMC)~\cite{duane1987hybrid,betancourt2017conceptual}, consisting of Markov chain Monte Carlo sampling. The HMC implementation in \textsc{Abinit} was used, considering a time step of 0.7 fs and 5000 steps in total (with 40 Monte Carlo sweeps between each molecular dynamics step). The determination of the critical temperature was carried out using a supercell of $6$ u.c. along the $z$-direction as this parameter seems to affect the numerical result, whereas for the computational experiments involving electric fields we used a supercell of $1$ u.c. along the $z$-direction to reduce computational cost.

In order to compute the polarization profile we followed the regular approach where local polarizations are computed within a linear approximation of the product of the Born effective charge tensor times the atomic displacements from the reference structure positions. An average over a volume of one unit cell centered at the Ti site was used. Therefore, the displacements taken into account involve the given Ti site, its six nearest oxygen neighbors and its eight nearest Ba neighbors. The resulting value was divided by the volume of the unit cell~\cite{meyer2002ab}. For KNbO$_\mathrm{3}$, the local polarization has been calculated using a similar approach and with the Born effective charge tensor from Ref.~\cite{wan2012structural}.

Finally, in order to account for the effect of the inhomogeneous electric fields an external force is added to each atom. Since the Born effective charge tensor ${\bold Z}^*$ describes the linear relation between the force on an atom and
the macroscopic electric field ${\bold E}$~\cite{Gonze-97}, the force induced along $\beta$ on an atom $\kappa$ at a given position ${\bf r}$ by a field in direction $\alpha$ is equal to $F_{\kappa,\beta}  = {Z}^*_{k,\alpha\beta}{E}_{\alpha}({\bf r})$.
\section{Skyrmions and antiskyrmions in KN\MakeLowercase{b}O\textsubscript{3}}
In this section, we present the DFT-based results of the relaxations of skyrmions and anti-skyrmions defects centered on K and Nb cations in analogy to those reported in the main text for the case of BaTiO$_3$. 
\begin{figure}[hbtp]
     \centering
      \includegraphics[width=8cm]{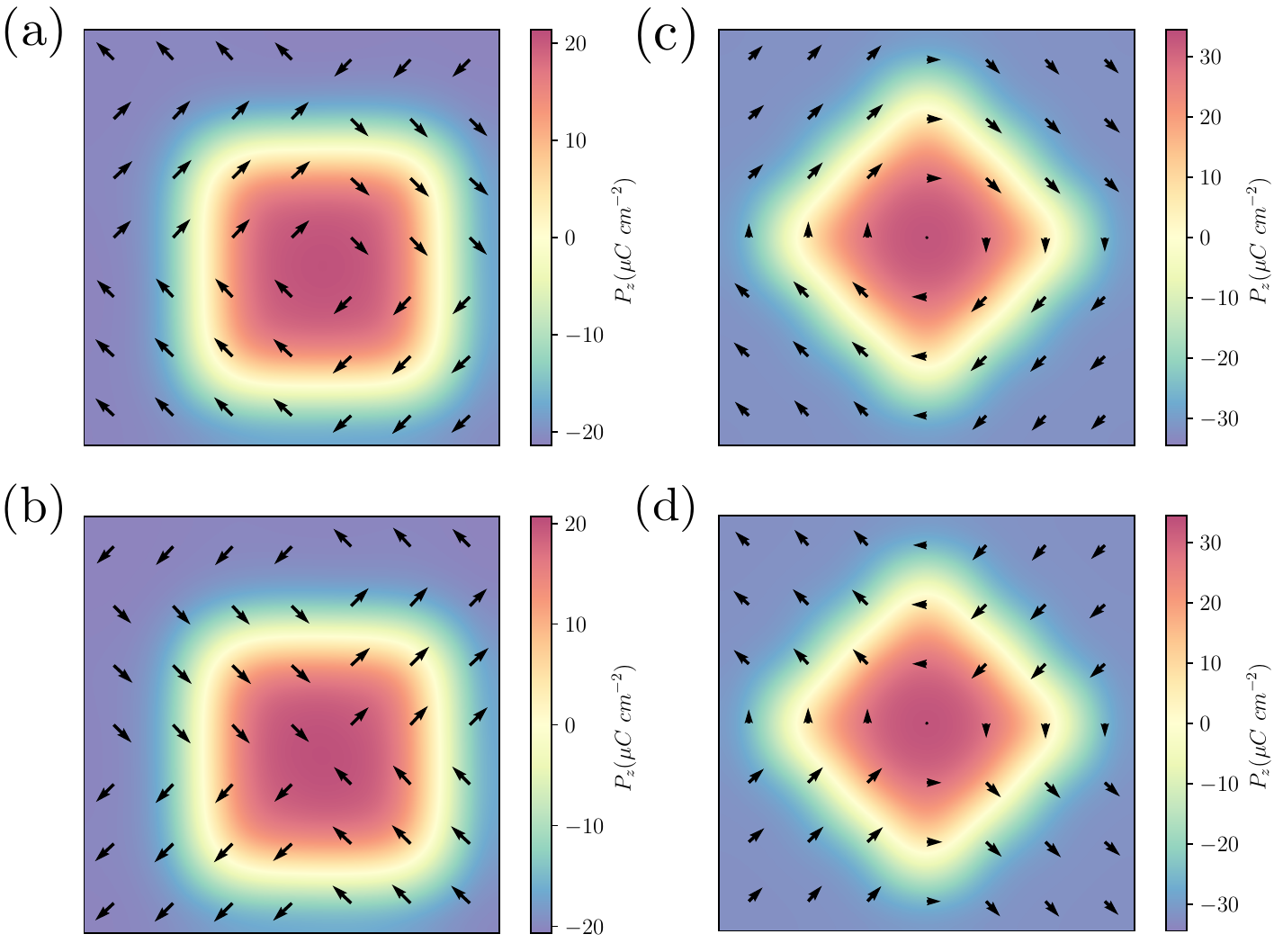}
      \caption{Polarization map with respect to the cubic reference and the unit cell centred on Nb atom. (a) K-centered skyrmion (b) K-centered anti-skyrmion (c) Nb-centered skyrmion (d) Nb-centered anti-skyrmion}
      \label{fig:Sk_aSk_KNbO3} 
\end{figure}

\begin{figure}[hbtp]
     \centering
      \includegraphics[width=5cm]{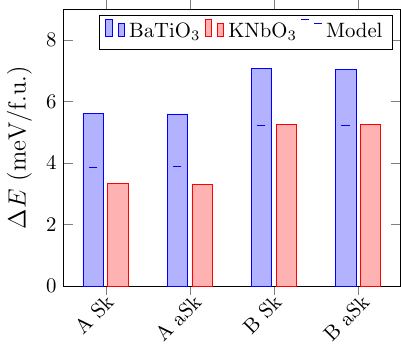}
      \caption{Energy comparison of skyrmion and anti-skyrmion defects centered on A and B sites for BaTiO$_\mathrm3$ (red) and KNbO$_\mathrm3$ (blue). DFT energies with respect to the rhombhohedral phase are represented by vertical bars whereas second-principles energies for the BaTiO$_\mathrm3$ model are represented by horizonatal lines within the blue bars.}
      \label{fig:Summary} 
  \end{figure}

As shown in Figure \ref{fig:Sk_aSk_KNbO3}, both skyrmion and anti-skyrmions centered on K and Nb cations can be stabilized at 0K being analogous to the ones found in BaTiO$_\mathrm{3}$ as shown in Figure \ref{fig:NEB}. Moreover, as illustrated in Figure \ref{fig:Summary}, skyrmions and anti-skyrmions centered on a given cation in both  BaTiO$_\mathrm{3}$ and KNbO$_\mathrm{3}$  are almost degenerated, with an energy difference of less than 0.04 meV/f.u. lying in the accuracy of DFT. Interestingly, as discussed on the main text, A-centered defects are energetically favored with respect to their B-centered counterparts. 

\section{Computation of the topological charge}
In this section, we shall compute the topological charge for the presented structures in the main body of the article.
The topological charge is computed by integrating the Pontryagin densitiy following Eq.~\ref{eq:top_charge} in transversal sections perpendicular to the direction of the tube
\begin{equation}
    \mathcal{Q}=\frac{1}{4\pi}\iint \vec{n}\cdot\left(\frac{\partial \vec{n}}{\partial x}\times\frac{\partial\vec{n}}{\partial y}\right)dxdy,
    \label{eq:top_charge}
\end{equation}
where $\vec{n}$ represents a normalized direction vector pointing along the polarization direction at each lattice site. For numerical stability we followed the recipe described in Ref.~\cite{Breg-81} to compute the integrand of Eq.~\ref{eq:top_charge} at each lattice site as it has been shown to perform better than a finite difference approach for rapidly evolving vector fields~\cite{Mauro-19}.

\begin{figure}[hbtp]
     \centering
      \includegraphics[width=12cm]{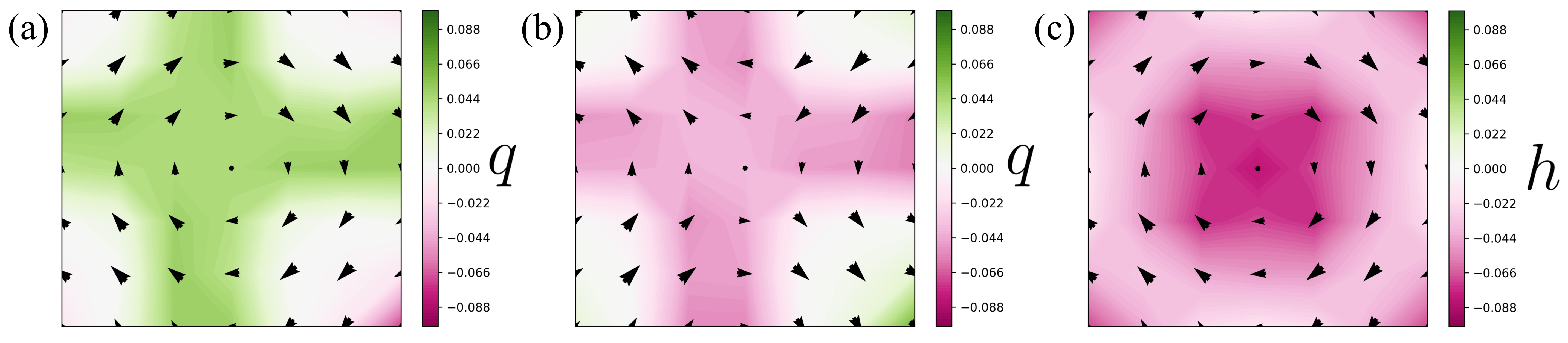}
      \caption{Analysis of the topological charge and helicity of the skyrmion tubes shown in Fig. 1 of the main body of the manuscript. Arrows indicate the direction of the in-plane polarization components. (a,b) The color map represents the Pontryagin density, integrand of Eq.~(\ref{eq:top_charge}) for the (a) skyrmion case and (b) antiskyrmion case. (c) The color map represents the helicity denisity computed as the integrand of Eq.~(\ref{eq:helicity}).}
      \label{fig:SP_Top_charge} 
  \end{figure}

As we show in Fig.~\ref{fig:SP_Top_charge}(a), for the Skyrmion texture reported inf Fig. 1(b) of the main body of the manuscript an integer value of $\mathcal{Q}=+1$ is obtained. Moreover, performing similar analysis on the antiskyrmion texture, Fig.~\ref{fig:SP_Top_charge}(b), evidence an integrated value of $\mathcal{Q}=-1$ for its topological charge.

Finally, the chirality in ferroelectric systems can be computed by means of the Helicity~\cite{Shafer-18,Junquera-23,Moffat-92,Moffatt-14} defined as
\begin{equation}
    \mathcal{H}=\iiint
    \mathbf{P}(\mathbf{r})\cdot\left(\nabla\times\mathbf{P}(\mathbf{r})\right)d\mathbf{r},
    \label{eq:helicity}
\end{equation}
where $\mathbf{P}(\mathbf{r})$ represents the local value of the polarization at the site $\mathbf{r}$. As shown in Fig.~\ref{fig:SP_Top_charge}(c), the chiral nature of the Skyrmion tube is determined by the coupling between the polarization rotation defined by the vortex's rotational direction and the polarization value of the polar nanodomain at the core of such vortex.

\section{B\MakeLowercase{a} and T\MakeLowercase{i} centered defects}
\begin{figure}[hbtp]
     \centering
      \includegraphics[width=10cm]{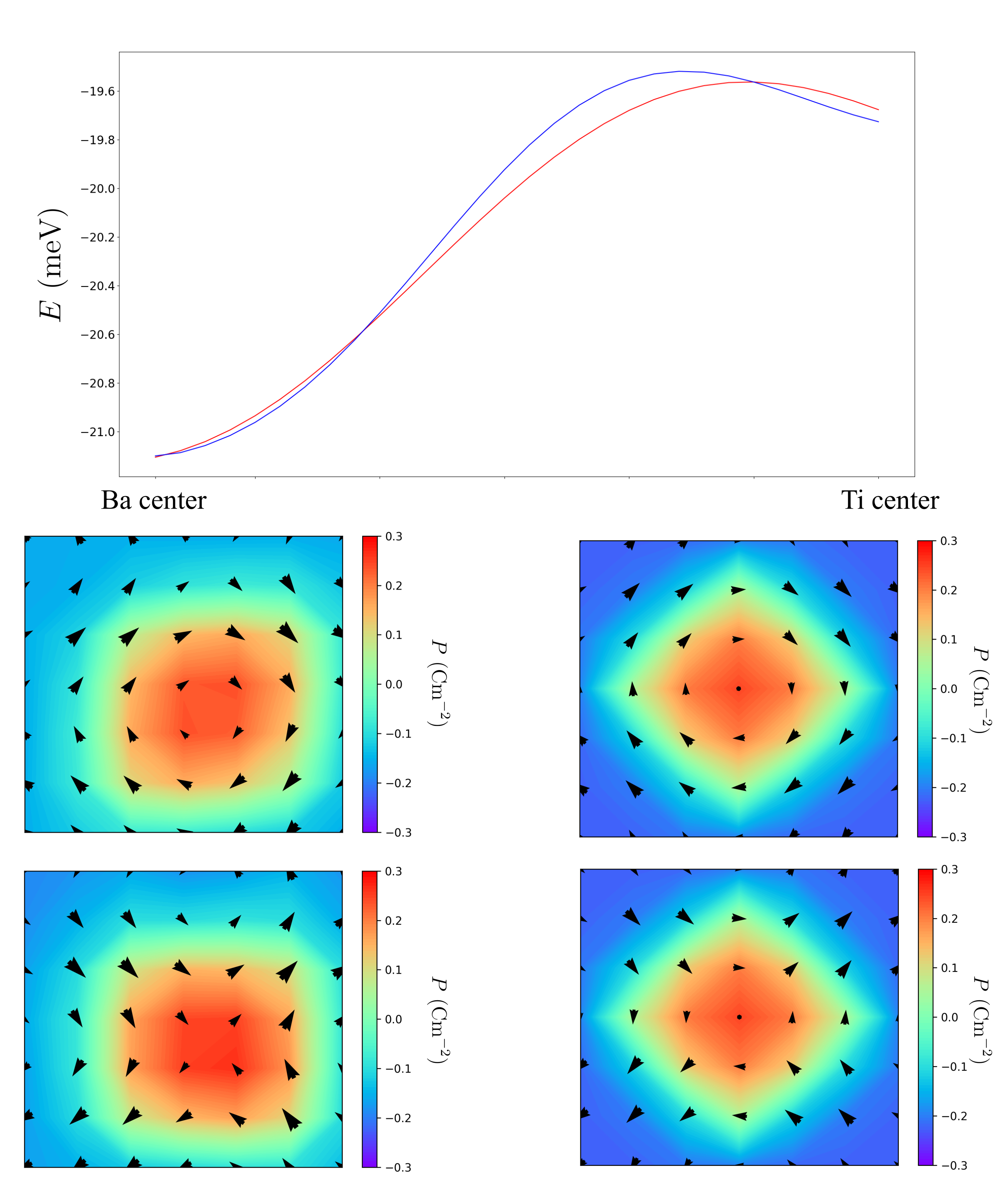}
      \caption{Energy barrier for the Ba to Ti centered skyrmions and antiskyrmions obtained from a NEB calculation with the second-principles model. Blue and red curves represent the energy per formula unit with respect to the cubic reference for antiskyrmions and skyrmions respectively. Below we can observe the dipole arrangements computed at the Ti site where arrows indicate the in-plane components and the colorbar represents the out-of-plane component.}
      \label{fig:NEB} 
  \end{figure}

In this section, we present the energy differences between Ba-centered and Ti-centered defects. As noted in the main text, both defect structures are stable at the DFT and second-principles levels, with an energy difference of approximately 1.2 meV per formula unit for our 7$\times$7$\times$1 supercell. Figure~\ref{fig:NEB} illustrates the transition pathway between Ba- and Ti-centered defects, derived from a nudged elastic band (NEB)~\cite{Jonsson-98,henkelman2000improved,sheppard2008optimization} calculation using our second-principles model. Throughout the transition path, the energies of the skyrmion and antiskyrmion textures remain nearly degenerate, indicating that the energy landscape is comparable for both textures in either defect configuration. As the NEB calculation reveals, the configuration where the in-plane vortex or antivortex singularity is positioned at a Ba site (left side of the figure) is energetically favored. This energy difference arises from the energy disparity between Ba-centered and Ti-centered domain walls, as the singularity of the in-plane components corresponds to the intersection of the domain walls separating the up/down and left/right domains.
\section{Asymptotic behavior of the defect's energy with increasing supercell size}
\begin{figure}[ht]
     \centering
      \includegraphics[width=7cm]{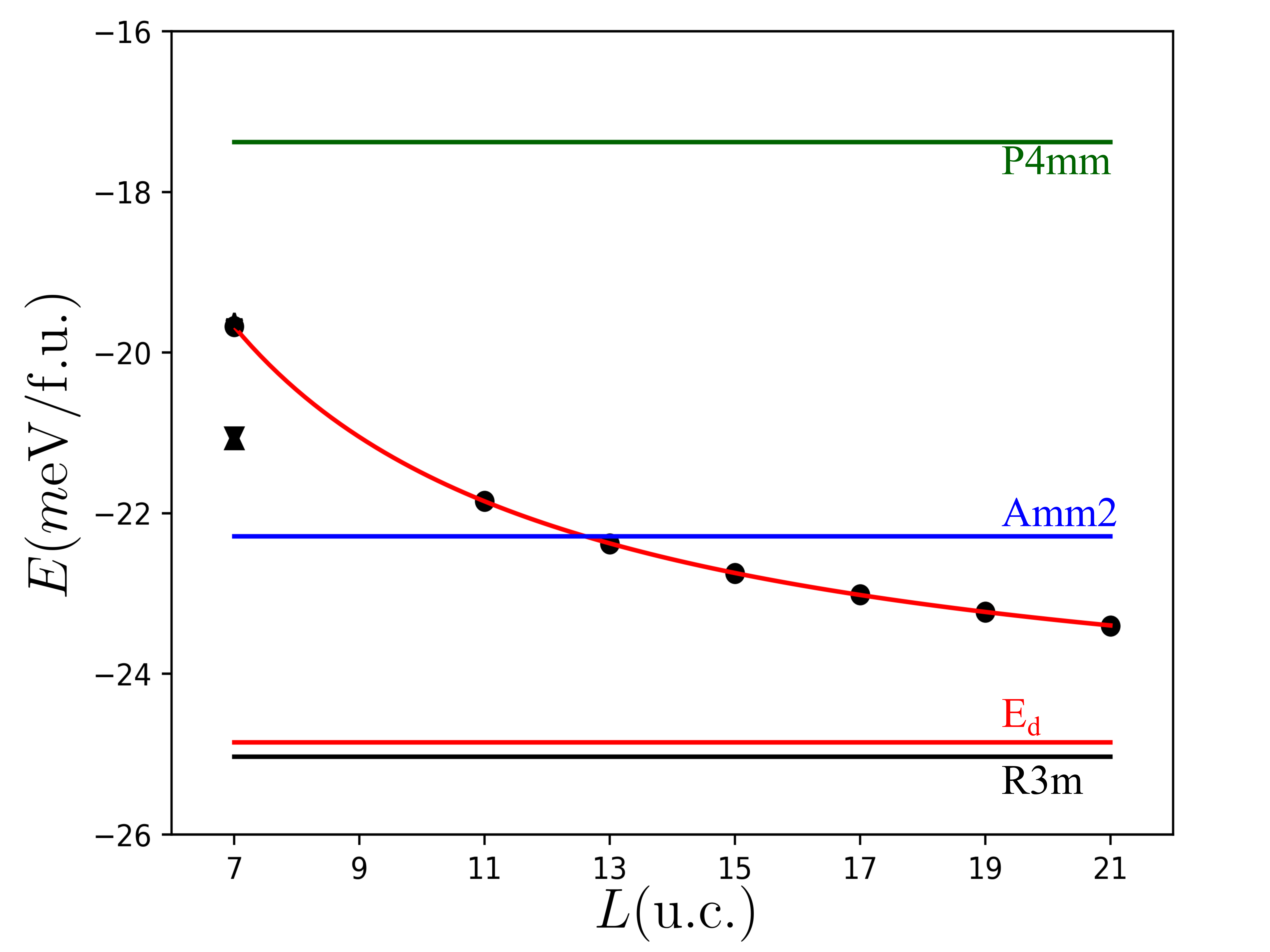}
      \caption{Energy dependence of a Ti-centered skyrmion tube exhibiting a nanocolumn of 2 u.c. as a function of supercell size, obtained from second-principles simulations. Black points indicate the simulation data, while the red line represent a fit to the power-law model and its asymptotic value as explained in the text. The black, blue and green lines represent the energy of the rhombohedral, orthorhombic and tetragonal structures respectively. Black star represents the value of the Ti-centered antiskyrmion whereas up/down triangles represent the Ba-centered skyrmion and antiskyrmion respectively. All energies are given with respect to the cubic reference.}
      \label{fig:asymptote} 
  \end{figure}

In this section we will analyze the asymptotic behavior of the Ti-centered skyrmion tube energy computed with second-principles as we increase the size of the simulation supercell. As explained in the main text, keeping a fixed value of two unit cells for the nanocolumn, we increase the value of the supercell in which the defect is embedded i.e. $L=7,~11,~13,~15,~17,~19,~21$. Even though the size of the nanocolumn is fixed, the in-plane components of the polarization extend throughout the whole supercell as discussed in the main text, in order to locally resemble the nearest $\langle 111\rangle$ R3m configuration. Therefore, as it can be observed in Fig.~\ref{fig:asymptote} an asymptotic behavior of the energy as we increse the supercell size is obtained. 

After fitting the second-principles energy to a power law model $E=E_{d}+a/(x+b)$ (red line) we observe that the model perfectly explains the data. The value of $E_d$ corresponds to the energy of the disturbed field in the limit of an infinite supercell and is represented by a red horizontal line in Fig.~\ref{fig:asymptote}. As it can be observed, the value of $E_d = -24.87 \pm 0.01$ meV/f.u. with respect to the cubic reference, is remarkably close to that of the rhombohedral monodomain phase, black curve in Fig.~\ref{fig:asymptote} ($E_{\rm{R3m}} = -24.94$ meV/f.u.), and lower than the tetragonal ($E_{\rm Amm2} = -17.38$ meV/f.u.) and the orthorhombic ($E_{\rm Amm2} = -22.29$ meV/f.u.) monodomain phases, green and blue curves in Fig.~\ref{fig:asymptote} respectively. For completeness, the energies of the reported Ba-centered and Ti-centered skyrmions and antiskyrmions in the 7$\times$7$\times$1 supercell are included in the figure as explained in the caption.
\section{Stability of the nanocolumn}
\begin{figure}[hbtp]
     \centering
      \includegraphics[width=10cm]{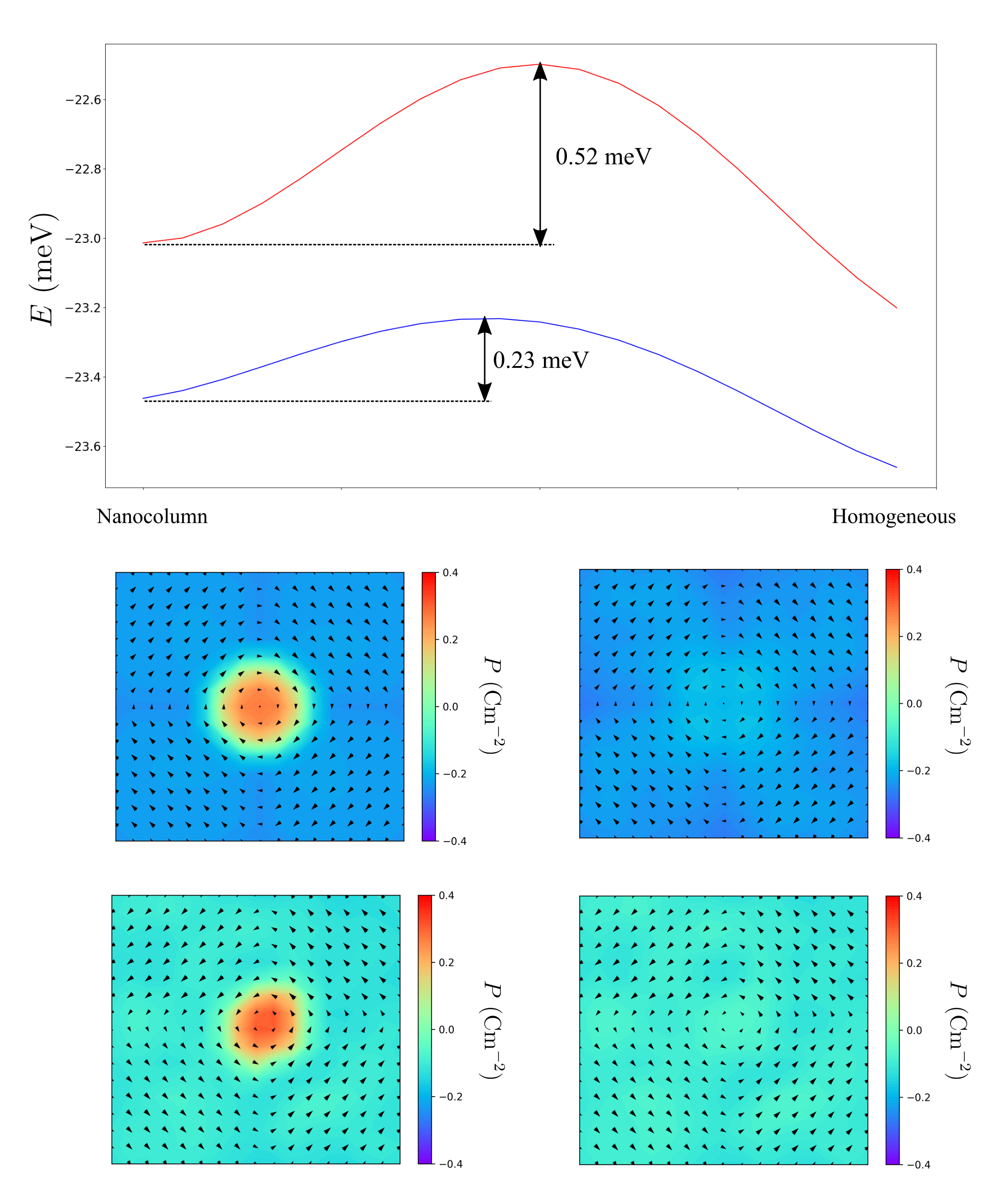}
      \caption{Energy barrier for the stability of Ba and Ti centered skyrmions towards an homogeneous out-of-plane polarization state obtained from a NEB calculation with the second-principles model. Blue and red curves represent the energy per formula unit with respect to the cubic reference for Ba and Ti centered defects respectively. In the dipole arrangements, we can observe the initial and final states where arrows indicate the in-plane components of the polarization and the colorbar represents the out-of-plane component.}
      \label{fig:NEB_nanocol} 
  \end{figure}

In this section, we examine the stability of the out-of-plane polar nanodomain as a function of whether the defect is Ba-centered or Ti-centered. As discussed in the main text, although the Ba-centered defect is energetically more favorable, it demonstrates lower thermal stability compared to the Ti-centered defect. Our second-principles molecular dynamics simulations indicate that the disappearance of the skyrmionic state arises from the destabilization of the polar nanocolumn, which transitions into a homogeneous out-of-plane polarization state. Here, we perform NEB calculations~\cite{Jonsson-98} to quantify these energy barriers. In Fig.~\ref{fig:NEB_nanocol}, we report the transition path of the disapearance of the polar nanocolumn for Ba-centered (blue curve and bottom dipole panels) and Ti-centered (red and top dipole panels) defects. As expected, the energy of the Ba-centered defects is lower than that of the Ti-centered ones; however, the energy barrier for the transition to a homogeneous out-of-plane polarization component is unexpectedly reduced by half. This result aligns well with the critical temperatures of $T=20$ K and $T=50$ K reported in the main text. In supplemental videos 2 and 3 we illustrate the evolution of the polarization pattern along the transition path reported in Fig.~\ref{fig:NEB_nanocol} where the disappearance of the nanocolumn in favor of a homogeneous out-of-plane polarization state can be observed.  Remarkably, the in-plane projection of the polarization remain unchanged throughout the process.
\section{Stabilization of Antivortex Textures via Contiguous Electrode Activation}
\begin{figure}[hbtp]
     \centering
      \includegraphics[width=9cm]{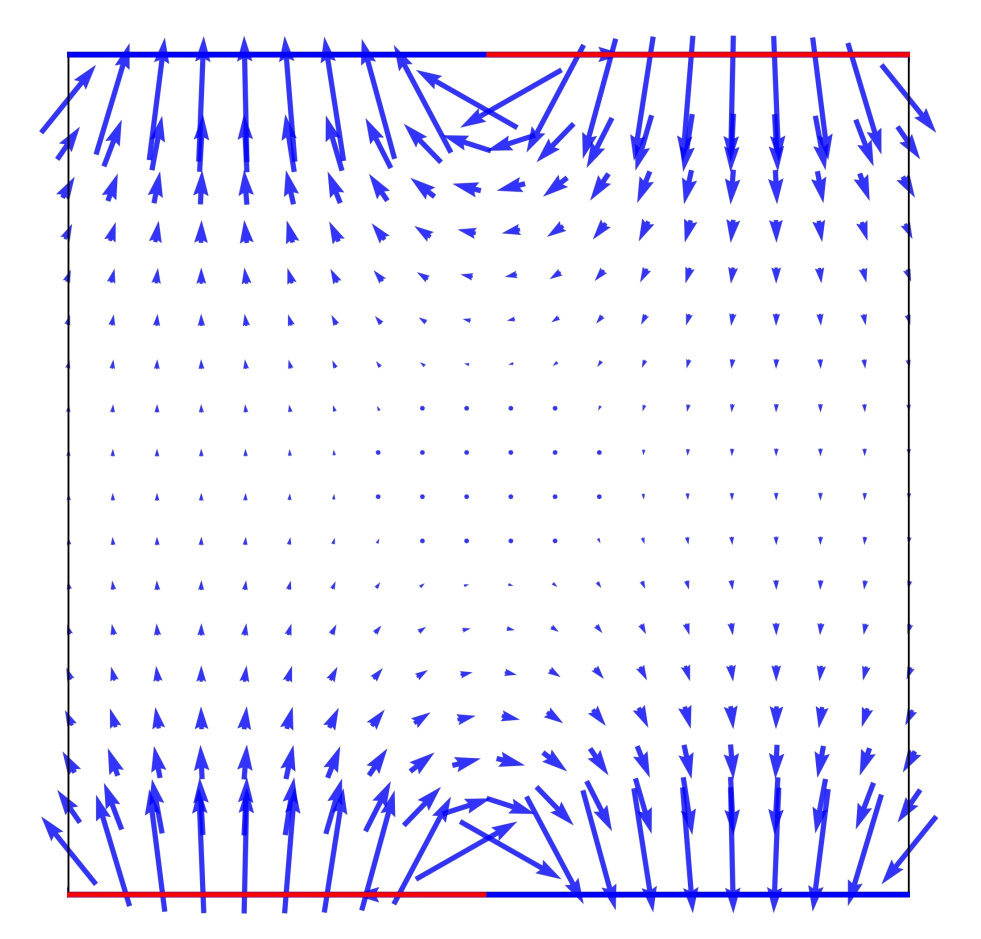}
      \caption{Schematic illustration of two contiguous electrodes with opposite polarity and the corresponding computed electric field generated when they are activated simultaneously. Blue arrows indicate the direction of the electric field following an antivortex pattern.}
      \label{fig:electrodes} 
  \end{figure}

In Fig.~\ref{fig:electrodes} we present a schematic illustration of the electrodes set up discussed on the main text in Fig. 3(a). As discussed in the main body of the paper, when activated sequentially in a stepwise manner they will induce up and down polarization domains within the material. However, if they are activated simultaniously edge effects between contiguous electrodes change the shape of the electric field.
The figure highlights the spatial distribution of this electric field, showcasing the strong in-plane components between the electrodes. 
As it is evident from the figure the shape of the electric field clearly follows an antivortex pattern similar to the in-plane polarization arrangements that the BaTiO$_3$ adopts in the antiskyrmion texture. 
This setupt demonstrates that the ability to experimentally generate an electric field with an antivortex-like pattern might indeed be feasible.
\medskip
\end{document}